\documentclass[twoside,a4paper,11pt]{sea9}
\usepackage{graphicx}
\usepackage{hyperref}
\usepackage{movie15}
\begin{document}
\pagenumbering{arabic}
\pagestyle{myheadings}
\thispagestyle{empty}
{\flushleft\includegraphics[width=\textwidth,bb=58 650 590 680]{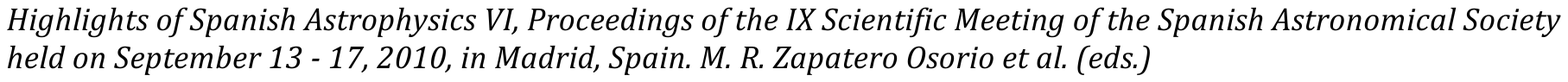}}
\vspace*{0.2cm}
\begin{flushleft}
{\bf {\LARGE
%
Gamma rays from extragalactic astrophysical sources
%
}\\
\vspace*{1cm}
%
Valent\'i Bosch-Ramon$^{1}$
%
}\\
\vspace*{0.5cm}
%
$^{1}$ Dublin Institute for Advanced Studies, 31 Fitzwilliam Place, Dublin 2, Ireland.\\ 
E-mail: valenti@cp.dias.ie
\\
%
\end{flushleft}
%
\markboth{
Gamma rays from extragalactic sources
}{ 
%
V. Bosch-Ramon
%
}
\thispagestyle{empty}
\vspace*{0.4cm}
\begin{minipage}[l]{0.09\textwidth}
\ 
\end{minipage}
\begin{minipage}[r]{0.9\textwidth}
\vspace{1cm}
\section*{Abstract}{\small
%
Presently there are several classes of detected gamma-ray extragalatic 
sources. They are mostly associated to active galactic nuclei (AGN) and (at soft gamma rays) 
to gamma-ray bursts (GRB), but not only.
Active galactic nuclei consist of accreting supermassive black holes
hosted by a galaxy that present in some cases powerful relativistic jet activity. These
sources, which have been studied in gamma rays for several decades, are
probably the most energetic astrophysical objects, and their appearance
depends much on whether their jets point to us. 
Gamma-ray bursts, 
thought to be associated to collapsing or merging stellar-mass objects at
cosmological distances, are also accreting highly relativistic 
jet sources that shine strongly at high energies. 
These are very short-duration events, but they are also the 
most luminous. 
Recently, star formation galaxies have turned out to be also 
gamma-ray emitters.
On the other hand, clusters of galaxies have not been 
detected beyond X-rays yet. These are the largest known
structures in the Universe; in their formation through accretion and
merging, shocks and turbulence are generated, which may lead to gamma-ray production. In this
work, the gamma-ray physics of AGNs is briefly presented, as well as that of starburst 
galaxies, GRBs 
and clusters of galaxies.
Afterwards, we consider some particular cases of
gamma-ray production in non-blazar AGN jets interacting with their medium at different scales.
%
\normalsize}
\end{minipage}
%
%
%
\section{Introduction \label{intro}}

Some extragalactic objects are known to be gamma-ray emitters since several decades ago. In particular, blazars, active galactic nuclei (AGN) with the jet pointing towards the observer
\cite{begelman84}, were the first to be discovered (e.g. \cite{swanenburg78}). However, given the late development of gamma-ray astronomy compared with other wavelengths, less technically
demanding, for a long time blazars were the only extragalactic source class clearly identified in hard gamma rays (a discussion on the first EGRET results can be found in \cite{dermer92}). 

In the EGRET era (1991--2000), there was already some evidence that non-blazar (misaligned) AGNs \cite{hartman99,sre99} and gamma-ray bursts (GRB)  \cite{hurley94} were hard gamma-ray
emitters, but the situation of these sources was far less clear than in the case of blazars. It has been with the present generation of satellite-borne (GeV; {\it AGILE} and {\it Fermi})
and ground based Cherenkov (TeV; HESS, MAGIC I-II and VERITAS) telescopes, that the field of extragalactic gamma-ray astronomy has become more crowded and its source classes better defined.
Presently, beside blazars, misaligned AGNs have been clearly established as a gamma-ray emitting class \cite{abdo10a}, together with GRBs (see, e.g. \cite{ghisellini10a} and references
therein)  and starburst galaxies \cite{acciari09,acero09,abdo10b}. Clusters of galaxies, although a well established type of non-thermal source, still remain undetected in the GeV and
TeV energies \cite{ackermann10,aharonian09,aleksic10}. 

In this paper, the main aspects of gamma-ray emission in AGNs and other extragalactic sources are briefly summarized. Then, the focus is put on some scenarios of gamma-ray production at
different spatial scales in the jets of non-blazar AGNs. The topic of gamma rays of dark matter origin is not treated here, for which the reader is referred to the literature (e.g.
\cite{bertone05}).

\section{Gamma-ray emission in extragalactic sources \label{extr}}

From a general point of view, the gamma-ray production in extragalactic astrophysical sources shares much in common with that in galactic ones, like supernova remnants, pulsar wind nebulae,
binary systems, star forming regions, and the galactic center. The main ingredients are collisionless shocks or turbulent magnetized plasmas embedded in dense radiation and/or matter
fields. Particles, electrons and/or protons/nuclei are accelerated in these environments and interact with other particles, either photons or nuclei, to produce gamma rays through inverse
Compton (IC), relativistic bremsstrahlung, proton-proton interactions, photomeson production, or photodisintegration \cite{blumenthal70,kelner06,kelner08,anchordoqui07}. Unlike galactic
sources, however, once created, gamma rays of extragalactic origin have to cover large distances and are subject to absorption through pair creation when interacting with the extragalactic
background radiation \cite{gould67}. This effect starts to be important for sources located significantly beyond $\sim 100$~Mpc.

\subsection{AGNs: blazars and misaligned jet sources}

From the point of view of the radio emission, and the non-thermal activity in general, there are two types of AGNs: radio quiet and radio loud \cite{antonucci93}. The latter sources present
strong emission from radio to gamma rays, and can be classified to their turn depending on the orientation of their jets. As mentioned in Sect.~\ref{intro}, blazars have their jets pointing
towards us, and their dominant emission is that produced in the inner regions of the jet. This radiation is thought to suffer strong beaming through Doppler boosting in the direction to the
observer given the inferred relativistic velocities of the jet flow (see \cite{begelman84} and references therein). Apparent superluminal motion is a characteristic feature of radio
emission in blazars, which implies real velocities close to $c$ (e.g. \cite{pearson81}). Relativistic velocities has been also invoked, among other reasons, to explain why gamma rays are
not suppressed due to pair creation in the emitter itself (e.g. \cite{maraschi92}). On the other hand, there are radio loud AGNs, also detected in gamma rays, whose jets are misaligned with
the observer direction, and therefore their radiation is not expected to be enhanced by Doppler boosting. In fact, unless they were very close to us, one would not expect that jet emission
could be detected in very misaligned jets, since Doppler (de)boosting actually would reduce the observed fluxes. This can be easily seen from this approximate relation for the radiation
luminosity enhancement for a moving blob: $L^{\rm obs}=\delta^4\,L^{\rm iso}$, where $\delta=1/\Gamma(1-\beta\cos\theta)$, $L^{\rm obs}$ is the observed luminosity around the spectral
energy distribution  maximum, $L^{\rm iso}$ the corresponding luminosity in the jet reference frame, $\Gamma$ the Lorentz factor of the blob, $\beta=(1-1/\Gamma^2)^{-1/2}$, and $\theta$ 
the angle between the velocity vector and the line of sight. For, say, $\Gamma=10$ (i.e. $\beta\approx 0.995$) and $\theta=5^\circ$, $L^{\rm obs}$ is $\approx 1.7\times 10^4\,L^{\rm iso}$,
whereas for $\theta=30^\circ$, $L^{\rm obs}$ is already just $\approx 0.27\,L^{\rm iso}$. This already shows, on one hand, the relevance of Doppler boosting when interpreting blazar
observations. In blazars, high Doppler boosting is actually the only way to reconcile observations with the available energetic budget (because of the large distances involved), and can
also explain the fast variability observed in these sources, since the observer time becomes shorter (or longer) by a factor $\delta^{-1}$. On the other hand, it is also worthy to note that
for misaligned sources one may be seeing very different radiation components than those of blazars. Components that are masked in the latter objects by strong inner-jet beamed emission may
be dominant when $\theta$ is large, and the radiation may be detectable if they are at short enough distances. 

The standard scenario for the gamma-ray emission in blazars, detected at GeV and TeV, is based on IC scattering. The IC process takes place either with synchrotron photons produced in the
same or nearby regions (synchrotron self-Compton), or with external photons generated in the accretion disk or the broad-line region (BLR) for bright nucleus AGNs \cite{bottcher07}.
However, it worthy noting that for this scenario to work, magnetic fields must be typically well below equipartition with matter. Low-energy  photons embedding the emitter, either of
internal or external origin, can absorb gamma rays reducing the observable fluxes (e.g. \cite{blandford95}; see also \cite{aha08} for the impact on spectrum formation). Also, gamma rays
absorbed in the surroundings of the source may lead to the generation of the so-called pair halos (e.g. \cite{aha94,kel04}). 

Particle acceleration in AGN jets may occur in shocks between different velocity jet shells close to the compact object, say within pc scales \cite{rees78,blandford79}. Magnetic
reconnection, taking place in plasmas with strong magnetic fields with near lines of different polarity, could also lead to relativistic particle production (e.g. \cite{larrabee03}). For
non-blazar or misaligned AGNs the gamma-ray emitter is not so well defined, existing several scenarios behind the gamma-ray emission in these sources, based either on leptonic or hadronic
processes (e.g. \cite{sta06,rieger08,lenain08,araudo10,barkov10,bordas10}). In Section~\ref{agn} below, some of these scenarios are presented in more detail. 

Beside shocks, second order Fermi (stochastic) and shear acceleration could also accelerate particles, in particular in large-scale jets \cite{rieger07}. It is worthy noting that when
modeling gamma-ray emission from AGNs, hadronic models, although not as successful in general as IC scattering, can be still feasible under certain conditions (e.g.
\cite{aharonian02,mucke03}).

\subsection{Starburst Galaxies}

Starburst galaxies present rates of star formation much higher than in our galaxy. Gamma-ray emission from these sources had been predicted in the past, since it is expected that a higher
formation rate will be linked to a higher supernova rate, and therefore to a high cosmic ray (CR) density. This, together to a higher density medium, leads to efficient gamma-ray production
through proton-proton interactions \cite{voelk89,voelk96,romero03}. This scenario is favored by the fact that there seems to be a correlation between the gamma-ray luminosity, the gas mass
and the supernova rate, as shown in \cite{abdo10b}. As noted in Sect.~\ref{intro}, these sources have been detected in both GeV \cite{abdo10b} and TeV \cite{acciari09,acero09} energy bands.

\subsection{GBRs}

After their discovery in the sixties of the last century by the VELA satellites (see \cite{klebesadel73}), the origin of GRBs remained unknown till the nineties, when BATSE and {\it
Beppo-Sax} unveiled their isotropic distribution in the sky as well as their redshifts \cite{fenimore93,metzger97}. The huge amounts of energy involved in GRBs, whicht allow detections at
very high redshifts, and their very fast variability, lead to the suggestions that the radiation was very strongly beamed, probably due to Doppler boosting. Also, the compactness of the
emitter implied by the variability timescales required very high Lorentz factors to avoid strong photon-photon absorption, as in the case of blazars. 

The non-thermal emission from GRBs is believed to be produced in two different phases. In the first phase, shells of different velocities collide producing the so-called prompt emission,
whereas in the second phase, the emission is due to a shock with the external medium (e.g. see \cite{meszaros06} and references therein). Gamma-ray bursts have been classified into two
categories depending on their duration: short vs long \cite{kouveliotou93}. The former are thought to be produced in the merger of two compact objects, whereas the latter are thought to
originate in the collapse of very massive stars, associated sometimes to observable supernova events \cite{galama98,kulkarni98,cas01}. Although our knowledge on GRBs has increased a lot
from the nineties when GeV emission was found in these objects, it is still unclear what is behind these radiation, whether IC or hadronic processes, although leptonic processes seem to be
favored \cite{dermer10}. It is also not clear in which phase the GeV emission is produced, either during the prompt emission phase or during the early-afterglow one (e.g.
\cite{wang09,ghisellini10}).

\subsection{Clusters of Galaxies}

Clusters of galaxies are the most recently virialized structures formed in the Universe. They contain galaxies, some of them AGNs, and keep accreting material from very large filaments of
matter, which are also part of the structure of the Universe at very large scales. Merging of clusters also take place, forming larger objects of the same kind. Accretion and cluster
merging lead to shocks and to the generation of turbulence, both suitable situations for particle acceleration and, possibly, gamma-ray emission. Also, hosted galaxies can pump non-thermal
plasma into the cluster. 

At present, only radio emission has been clearly detected from clusters, and there is also some evidence of non-thermal radiation in the UV and the hard X-rays. Although expected
\cite{atoyan00,berrington03,araudo08,vannoni09}, these sources have not been detected yet in gamma rays, fact that allows to constrain the non-thermal contribution to the intracluster
medium pressure \cite{ackermann10,aharonian09,aleksic10}. Leptonic and hadronic mechanisms producing gamma rays have been postulated in clusters of galaxies, with different possible
emitting channels, IC and proton-proton collisions, and regions, the peripheral relics or the main cluster body \cite{pfrommer08}.

\section{Non-blazar AGN gamma-ray emission \label{agn}}

In this section some scenarios for the production of gamma rays in non-blazar AGNs are briefly discussed. This overview is not comprehensive but just focus on few possible mechanisms for
variable and steady, as well as point-like and extended, gamma-ray radiation in non-blazar AGNs.

\subsection{Transient and steady emission due to jet base/environment interactions}

The regions close to a galactic supermassive black hole (SMBH) are thought to be filled by a compact stellar population \cite{bis82}. It seems also likely that those regions may be
filled by inhomogeneous flows that could contain dense and cold clouds. These clouds would be photoionized by the radiation produced in the inner regions of the accretion disk surrounding
the SMBH, and could produce the ultraviolet (UV) lines with significant thermal broadening observed in powerful AGNs, the BLR \cite{kro81}. It has been proposed as well that, instead of
clouds, a stellar population, red giants (RG) for instance, may be behind the mentioned UV broad lines \cite{pens88}. In any case, both stars and clouds can interact with the jet in radio loud
AGNs leading to gamma-ray production. On one hand, RGs could penetrate the base of the jets in nearby AGNs, like Cen~A and M87, with subsequent lost of the external layers of the
star due to the jet ram/magnetic pressure. The lost matter is an effective obstacle for the jet flow and can tap part of the jet energy in the form of heat, kinetic energy and radiation.
These events are not expected to last long, given the interaction characteristics, and can be behind transient gamma-ray events in nearby misaligned jet sources \cite{max10}. If the
magnetic field in the base of the jet were strong, hadronic processes would be more suitable to generate gamma rays, since IC radiation would be strongly suppressed due to synchrotron
cooling. 

\begin{figure}
\center
\includegraphics[scale=0.3]{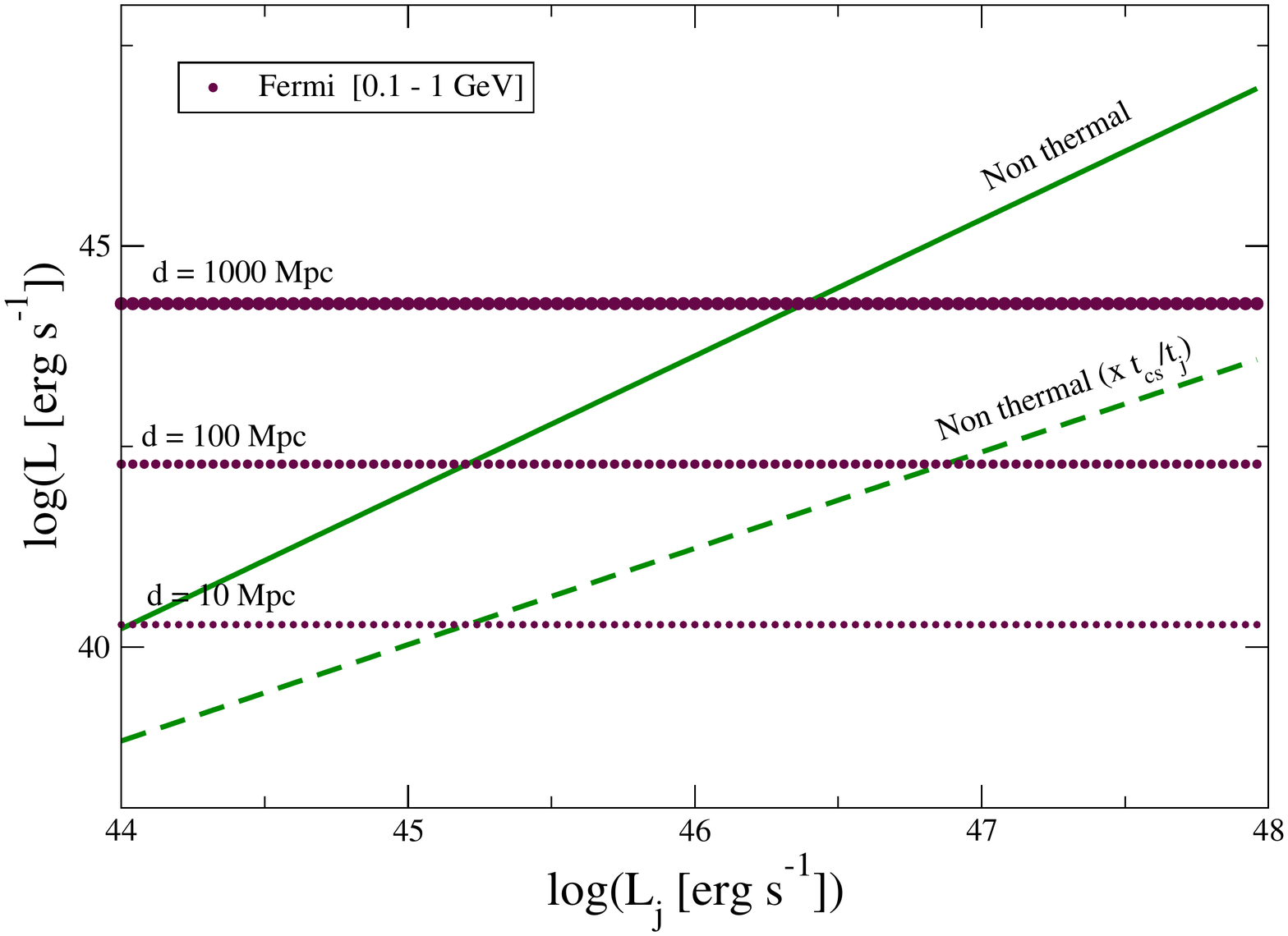} 
\caption{\label{fig1} Computed high-energy 
IC luminosity in the high-energy range for BLR clouds interacting with an FRII AGN jet for different jet kinetic luminosities. It is compared with the 
approximated {\it Fermi} sensitivity for a source located at three different distances: 10, 100 and 1000~Mpc. Two lifetimes of the clouds inside the jet are 
considered to calculate 
the luminosities, one is the jet crossing time, and the other is the time in which the whole cloud gets shocked by the jet. The real cloud lifetime is expected to be between these 
two timescales (adapted from \cite{ara10}).}
\end{figure}

Beside stars, clouds from the BLR\footnote{In fact, dark clouds may be also present even if the BLR region is not formed.} could interact as well with the jets leading to gamma-ray
emission. In this scenario, rare interactions of a big cloud with the jet base in low-luminous nearby AGNs may also, as in RG-jet interactions, lead to gamma-ray flares. In addition, for
powerful sources with a fully developed BLR, the expected penetration of the jet by many BLR clouds can yield a steady source of high-energy radiation \cite{ara10}. In all these models,
significant radio emission is not expected from the accelerator, since it is too compact and radio attenuation will be strong because of non-radiative losses and synchrotron self-absorption.
However, this radiation could be produced farther downstream in the jet, in relativistic electrons accelerated in the cloud shock and carried away by the postshock material. In
Fig.~\ref{fig1}, it is shown, for different distances, the approximated {\it Fermi} sensitivity together with the expected luminosity of the interaction of many clouds in bright non-blazar
AGNs.

\subsection{Emission from the termination regions of FRI jets}

\begin{figure}
\center
\includegraphics[scale=0.6]{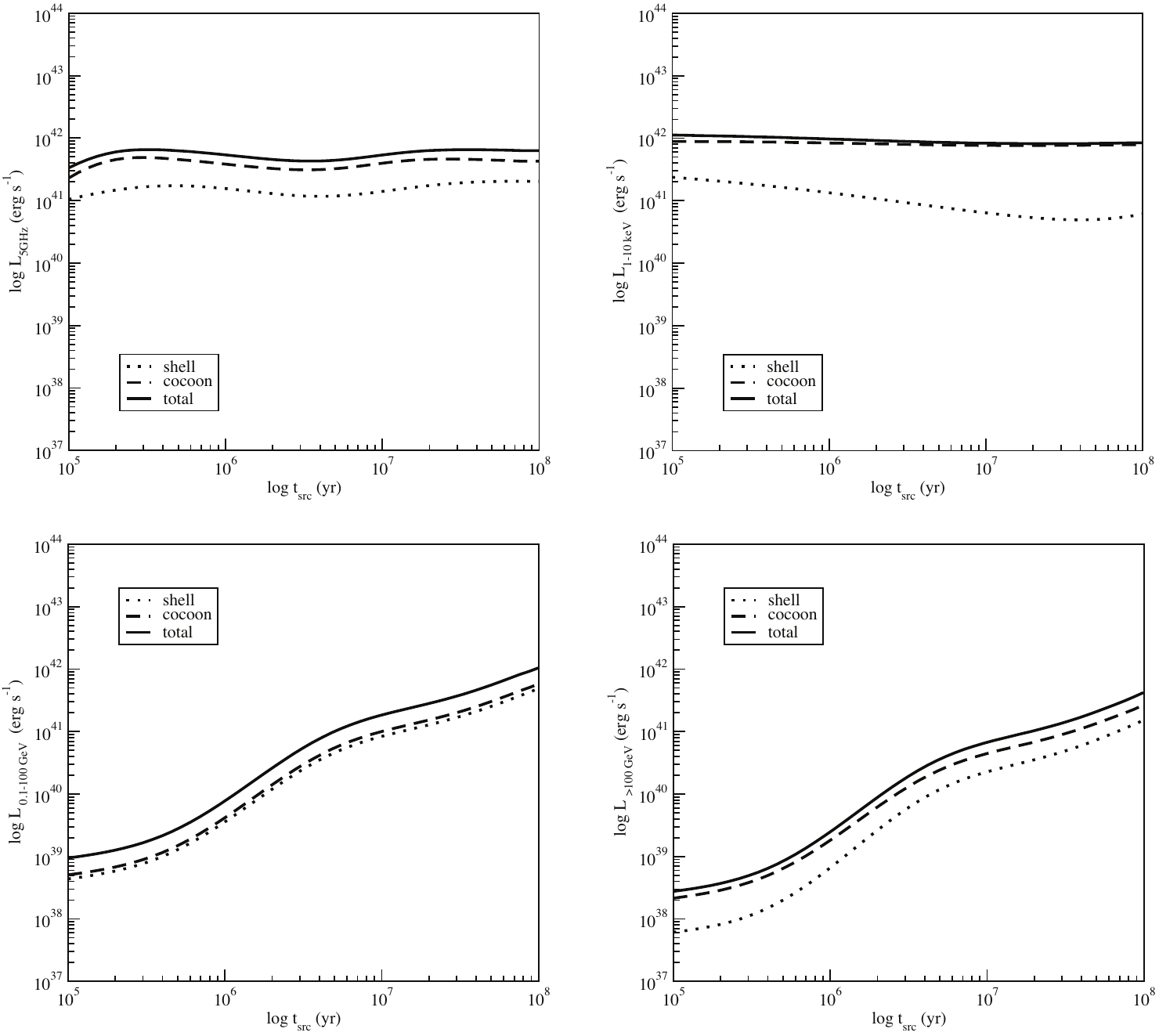} 
\caption{\label{fig2} Evolution of the computed non-thermal 
luminosities in radio (5~GHz), X-rays (1-10~keV), and gamma rays (0.1--100~GeV and $>$100~GeV) for the termination region of the jet of an FRI radio 
galaxy (adapted from \cite{bor10}).}
\end{figure}

Although the termination regions of AGN jets are well established non-thermal emitters, gamma-ray emission was not observed from these structures till recently, with the detection of the
Cen~A giant radio lobes by {\it Fermi} \cite{abdo10}. The gamma rays are extended and come from the largest radio lobes of this source. Cen~A is classified as a Fanaroff-Riley~I radiogalaxy
\cite{bick94}, and in these sources the jet is disrupted and much smaller than the whole jet/IGM structure (unlike FRII radiogalaxies; see \cite{fan74,per10}). However, strong shocks may be
linked to jet disruption, and relativistic particles may be injected in the disrupted jet medium by those shocks. The shock propagating in the external medium may be also able to accelerate
particles. For old enough sources, IC scattering off CMB photons becomes a dominant cooling channel for the accelerated particles, and if produced in sources at distances closer or $\sim
100$~Mpc, IC gamma rays could be detected by space- and ground-based instruments in GeV and TeV energies, respectively \cite{bor10}. In Fig.~\ref{fig2}, the predicted
radio-to-gamma-ray lightcurve for an FRI is presented, which shows the flux evolution with time.

\section{Final remarks}

The extragalactic field is nowadays much more populated by gamma-ray sources and classes of them than in the recent past, an improvement due to the last generation of GeV and TeV
instruments. In the same line, forthcoming instrumentation, like CTA in the TeV band, will allow a much better study of sources that are just detected at present, in particular misaligned
AGNs and starburst galaxies. GRBs may also present a very high-energy component, although for most of them a strong cutoff is expected around 100~GeV because of the cosmological distances
and the effects of the background absorbing radiation. The case of clusters of Galaxies is still unclear, but one expects that eventually a gamma-ray component will be detected, since it is
known that particle acceleration takes place in these objects, and there is no reason why particles should not reach gamma-ray producing energies. It is interesting to point out the
similarities in many aspects shared by the different extragalactic sources and the galactic ones. Beside their observational properties, like some are extended and others compact, there are
also similarities with galactic sources, like when debating the origin of the emission, leptonic vs hadronic, or the dominant acceleration mechanism. It is remarkable that shocks, jets,
turbulent flows, and magnetic fields play a role basically in all the astrophysical sources of gamma-rays. To finish, it is worthy noting the role of extragalactic gamma-ray sources in the
origin of ultra and extremely high-energy (UHE/EHE) CRs (e.g. \cite{ahar02}), since these sources are the best candidates to be the UHE/EHE CR accelerators.
%
\small  
%
\section*{Acknowledgments}   
%
The research leading to these results has received funding from the European Union
Seventh Framework Programme (FP7/2007-2013) under grant agreement
PIEF-GA-2009-252463
V.B-R.
acknowledges also support by the Ministerio de Educaci\'on y Ciencia (Spain) under grant AYA 2007-68034-C03-01, FEDER
funds.
%

%
\end{document}